\documentclass[aps,pra,showpacs]{revtex4}

\pdfoutput=0

\usepackage{graphicx}
\usepackage{graphicx}
\usepackage{bbm}
\usepackage{amsmath}
\usepackage{amsfonts}
\usepackage{amssymb}

\begin{document}

\title{Heisenberg's Uncertainty Relation and Bell Inequalities in High Energy Physics\\
\vspace{0.3cm}
\small{An effective formalism for unstable two-state systems}}
\thanks{Beatrix C. Hiesmayr wants to thank Gregor Weihs for the invitation to give a talk at the conference ``Advances in Quantum Theory'' (V\"axj\"o, Sweden, June 2010)
and Andrei Khrennikov for inviting her to organize a session ``Fundamentals of QM tested in High Energy Physics'' in the forthcoming conference
``Foundations of Probability and Physics-6'' (V\"axj\"o, Sweden, 13-16 June 2010), which initiated this work.}

\author{Antonio Di Domenico$^{1}$}
\email{antonio.didomenico@roma1.infn.it}

\author{Andreas Gabriel$^{2}$}
\email{Andreas.Gabriel@univie.ac.at}

\author{Beatrix C. Hiesmayr$^{2,3}$}
\email{Beatrix.Hiesmayr@univie.ac.at (Corresponding author)}


\author{Florian Hipp$^{2}$}
\email{Florian.Hipp@univie.ac.at}

\author{Marcus Huber$^{2}$}
\email{Marcus.Huber@univie.ac.at}

\author{Gerd Krizek$^{2}$}
\email{Gerd.Krizek@univie.ac.at}

\author{Karoline M\"uhlbacher$^{2}$}
\email{Karoline.Muehlbacher@univie.ac.at}
\author{Sasa Radic$^{2}$}
\email{Sasa.Radic@univie.ac.at}
\author{Christoph Spengler$^{2}$}
\email{Christoph.Spengler@univie.ac.at}
\author{Lukas Theussl$^{3}$}
\email{Lukas.Theussl@savba.sk}

\affiliation{$^{1}$Universit\`a degli Studi di Roma, La Sapienza, Piazzale Aldo Moro 5, 00185 Rome, Italy}
\affiliation{$^{2}$University of Vienna, Faculty of Physics, Boltzmanngasse 5, 1090 Vienna, Austria.}
\affiliation{$^{3}$Research Center for Quantum Information, Institute of Physics,
Slovak Academy of Sciences, D\`ubravsk\`a cesta 9, 84511 Bratislava, Slovakia}

\begin{abstract}
An effective formalism is developed to handle decaying two-state systems. Herewith, observables of such systems can be described by a single operator in the Heisenberg picture. This allows for using the usual framework in quantum information theory and, hence, to enlighten the quantum feature of such systems compared to non--decaying systems. We apply it to systems in high energy physics, i.e. to oscillating meson--antimeson systems. In particular, we discuss the entropic Heisenberg uncertainty relation for observables measured at different times at accelerator facilities including the effect of $\mathcal{CP}$ violation, i.e. the imbalance of matter and antimatter. An operator--form of Bell inequalities for systems in high energy physics is presented, i.e. a Bell--witness operator, which allows for simple analysis of unstable systems.
\end{abstract}

\maketitle

\section{Introduction}

The theoretical framework introduced in this paper can be applied in general to a broad variety unstable systems, however, the focus is on meson-antimeson systems  and their information theoretic interpretations of certain quantum features of single and bipartite (entangled) systems. In particular, we discuss meson-antimeson systems, e.g. the neutral K--meson or B--meson system, which are very suitable to discuss various quantum foundation issues (see e.g. Refs~\cite{HiesmayrKLOE,Mavromatos1,Mavromatos2,Giuseppe1,Blasone,Giuseppe2,Hiesmayr5,Genovese,Hiesmayr6,Genovese2,Caban,Hiesmayr9,Bramon3,Bramon4,Durt,Mavromatos3,LIQiao,Hiesmayr10,Yabsley,Bigi,DiDomenico,Capolupo1,Beuthe}). Neutral kaons are popular research objects in Particle Physics as they were the first system that was found to violate the $\mathcal{CP}$ symmetry (${\cal C}\dots$charge conjugation; ${\cal P}\dots$ parity), i.e. the imbalance of matter and antimatter. They are also well suited to investigate a possible violation of the $\mathcal{CPT}$ symmetry (${\cal T}\dots$ time reversal); see e.g. Refs.~\cite{DiDomenico,Capolupo1}.

Neutral meson--antimeson systems are oscillating and decaying two-state systems and can also described as bipartite entangled systems opening the unique possibility to test various aspects of quantum mechanics also for systems not consisting of ordinary matter and light.

The purpose of this paper is twofold. Firstly to enlighten that these systems provide different insights into quantum theory which are not available in other quantum systems via exploring e.g. the Heisenberg uncertainty relation in its entropic formulation or Bell inequalities which prove that there are correlations stronger than those obtainable in classical physics. Secondly, we introduce a comprehensive and simple mathematical framework which is close to the usual framework to handle stable systems and, therefore, allows for developing novel tools and potential applications.

In Section \ref{timeevolution} we introduce how the time evolution of neutral kaons are usually obtained. In Section \ref{AcceleratorFacilities} we discuss what kind of questions can be raised to the quantum system at accelerator facilities and what is measured at such facilities. In particular, we outline that there are two different measurement procedures not available to other quantum systems. Then the effective formulation of the observables corresponding to a certain question raised to the quantum system is introduced (Section \ref{effectiv}), which is our main result. Then we analyze different measurement settings and their uncertainty (Section \ref{Heisenberg}). In particular, we show that $\mathcal{CP}$ violation introduces an uncertainty in the observables of the mass eigenstates and, herewith, in the dynamics. Last but not least we proceed to bipartite entangled systems and present the generalized Bell--CHSH inequality for meson--antimesons systems~\cite{Hiesmayr3} in a witness form (Section \ref{BI}). This allows to derive the maximal and minimal bound of the Bell inequality by simply computing the eigenvalues of the effective Bell operator, i.e. without relaying on optimizations over all possible initial states.

\section{The Dynamics of Decaying and Oscillating Systems}\label{timeevolution}

The phenomenology of oscillation and decay of meson-antimeson systems can
be described by nonrelativistic quantum mechanics effectively, because the
dynamics are rather depending on the observable hadrons than on the more
fundamental quarks. A quantum field theoretical calculation showing negligible
corrections can e.g. be found in Refs.~\cite{Capolupo1,Beuthe}.

Neutral meson $M_0$ are bound states of quarks and antiquarks. As numerous experiments
have revealed the particle state $M_0$ and the antiparticle state $\bar M_0$ can decay into the same final
states, thus the system has to be handled as a two state system similar to
spin $\frac{1}{2}$ systems. In addition to being a decaying system these massive particles
show the phenomenon of flavor oscillation, i.e. an oscillation between matter
and antimatter occurs. If e.g. a neutral meson is produced at time $t = 0$ the
probability to find an antimeson at a later time is nonzero.

The most general time evolution for the two state system $M^0-\bar M^0$ including all its decays is given by an infinite--dimensional vector in Hilbert space
\begin{eqnarray}
|\tilde{\psi}(t)\rangle &=& a(t) |M^0\rangle+b(t) |\bar M^0\rangle+c(t)|f_1\rangle+d(t)|f_2\rangle+\dots
\end{eqnarray}
where $f_i$ denote all decay products and the state is a solution of the Schr\"odinger equation ($\hbar\equiv 1$)
\begin{eqnarray}
\frac{d}{dt}|\tilde{\psi(t)}\rangle&=&-i\hat{H}|\tilde{\psi(t)}\rangle\;
\end{eqnarray}
where $\hat{H}$ is an infinite-dimensional Hamiltonian operator. There is no method
known how to solve this infinite set of coupled differential equations affected
by strong dynamics. The usual procedure is based on restricting to
the time evolution of the components of the flavour eigenstates, $a(t)$ and $b(t)$. Then one uses the Wigner-Weisskopf approximation and can write down an effective Schr\"odinger equation
\begin{eqnarray}\label{schroedi}
\frac{d}{dt}|\psi(t)\rangle&=&-i\;H |\psi(t)\rangle\;
\end{eqnarray}
where $|\psi\rangle$ is a two dimensional state vector and $H$ is a non-hermitian Hamiltonian. Any non-hermitian Hamiltonian can be written as a sum of two hermitian operators $M,\Gamma$, i.e. $H=M+\frac{i}{2}\Gamma$, where $M$ is the mass-operator, covering the unitary part of the evolution and the operator $\Gamma$ describes the decay property.
The eigenvectors and eigenvalues of the effective Schr\"odinger equation, we denote by
\begin{eqnarray}
H\;|M_i\rangle &=&\lambda_i\; |M_i\rangle
\end{eqnarray}
with $\lambda_i=m_i+\frac{i}{2} \Gamma_i$. For neutral kaons the first solution (with the lower mass) is denoted by $K_S$, the short lived state, and the second eigenvector by $K_L$, the long lived state, as there is a huge difference between the two decay constants $\Gamma_S\simeq 600 \Gamma_L$.

Certainly, the state vector is not normalized for times $t>0$ due to the non-
hermitian part of the dynamics. Different strategies have been developed
to cope with that. We present here one which is based on the open quantum formalism, i.e. we show that the effect of decay is a kind of decoherence.

In quantum information theory and in experiments one often has to deal
with situations where the system under investigation unavoidable interacts with the environment which is in general inaccessible. In this case only
the joint system evolves according to the Schr\"odinger equation, it is unitary.
The dynamics of the system of interest then is given by ignoring all degrees
of freedom of the environment, by tracing them out. Such
systems are called open quantum systems and under certain assumptions they
may be described by a so called master equation.

In Ref.~\cite{BGH4} the authors showed that systems with non-hermitian Hamiltonians generally can be described by a master equation. As time evolves the kaon interacts with an environment which causes the decay. In our case the
environment plays the role as the QCD vacuum in quantum field
theory, but has not to be modeled, only the generators have to be defined describing the effect of the interaction. In particular the time evolution
of neutral kaons can be described by the master equation
(found by Lindblad \cite{Lindblad} and, independently, by Gorini, Kossakowski and Sudarshan \cite{GoriniKossakowskiSudarshan})
\begin{eqnarray}\label{masterequation}
\frac{d}{dt} \rho&=&-i [\cal{H},\rho]-\cal{D}[\rho]
\end{eqnarray}
where the dissipator under the assumption of complete positivity and
Markovian dynamics has the well known general form $ {\cal
D}[\rho]=\frac{1}{2}\sum_j ({\cal A}_j^\dagger{\cal
A}_j\rho+\rho{\cal A}_j^\dagger{\cal A}_j-2 {\cal A}_j\rho{\cal
A}_j^\dagger)$ with ${\cal A}_j$ are the generators. The density matrix $\rho$ lives on
$\textbf{H}_{tot}=\textbf{H}_s\bigoplus\textbf{H}_f$ where $s/f$
denotes ``surviving'' and ``decaying'' or ``final'' components,
and has the following decomposition
\begin{equation}\label{rhotot}
\rho=\left(\begin{array}{cc} \rho_{ss}&\rho_{sf}\\
\rho_{sf}^\dagger&\rho_{ff}\end{array}\right)
\end{equation}
where $\rho_{ij}$ with $i,j=s,f$ denote $2\times 2$ matrices. The
Hamiltonian $\cal{H}$ is the mass matrix $M$ of the effective
Hamiltonian $H$ extended to the total Hilbert space
$\textbf{H}_{tot}$ and $\Gamma$ of $H_{eff}$ defines a Lindblad
operator by $\Gamma=A^\dagger A$, i.e.
\begin{eqnarray*}
{\cal H}=\left(\begin{array}{cc} H&0\\
0&0\end{array}\right)\;,\; {\cal A}=\left(\begin{array}{cc} 0&0\\
A&0\end{array}\right)\quad\textrm{with}\quad A:
\textbf{H}_s\rightarrow \textbf{H}_f\,.
\end{eqnarray*}
Rewriting the master equation for $\rho$, Eq.~(\ref{rhotot}), on
$\textbf{H}_{tot}$
\begin{eqnarray}
\dot{\rho}_{ss}&=&-i[H,\rho_{ss}]-\frac{1}{2}\,\lbrace A^\dagger
A, \rho_{ss}\rbrace\;,\\
\dot{\rho}_{sf}&=&-i H \rho_{sf}-\frac{1}{2}\, A^\dagger
A\, \rho_{sf}\;,\\
\label{rhoff} \dot{\rho}_{ff}&=&A\,\rho_{ss}\,A^\dagger\,,
\end{eqnarray}
we notice that the master equation describes the original effective
Schr\"odinger equation (\ref{schroedi}) but with properly
normalized states (see Ref.~\cite{BGH4}). By construction the time
evolution of $\rho_{ss}$ is independent of $\rho_{sf}, \rho_{fs}$
and $\rho_{ff}$. Further $\rho_{sf},\rho_{fs}$ completely decouples
from $\rho_{ss}$ and thus can without loss of generality be chosen
to be zero since they are not physical and can never be measured. With
the initial condition $\rho_{ff}(0)=0$ the time evolution is
\textit{solely} determined by $\rho_{ss}$---as expected for a
spontaneous decay process---and formally given by integrating
the components of Eq.~(\ref{rhoff}). It proves that the decay is Markovian and
moreover completely positive.

Explicitly, the time evolution of a neutral kaon is given in the
lifetime basis, $\{K_S, K_L\}$, by ($$\rho_{ij}=\langle K_i|\rho| K_j\rangle$$, $\rho_{SS}+\rho_{LL}=1$):
\begin{eqnarray}\label{densitysingle}
\rho(t)=\left(\begin{array}{cccc} e^{-\Gamma_S t} \rho_{SS}& e^{-i
\Delta m t-\Gamma t} \rho_{SL}&0&0\\
e^{i \Delta m t-\Gamma t} \rho_{SL}^*&e^{-\Gamma_L t} \rho_{LL}&0&0\\
0&0&(1-e^{-\Gamma_L t}) \rho_{LL}&0\\
0&0&0&(1-e^{-\Gamma_S t}) \rho_{SS}
\end{array}\right)\,.
\end{eqnarray}
Note that, formally, one also obtains off--diagonal contributions in the $\rho_{ff}$ component, but as they cannot be measured we set them to zero without loss of generality.

The extension to bipartite systems is straightforward, i.e. by
\begin{eqnarray}
\cal{H}&\longrightarrow& \cal{H}\otimes\mathbbm{1}+\mathbbm{1}\otimes\cal{H}\nonumber\\
A_0&\longrightarrow&A_0\otimes\mathbbm{1}+\mathbbm{1}\otimes A_0
\end{eqnarray}
but we will not need to use that as our introduced effective formalism for single particles (Section \ref{effectiv}) generalizes simply for any multipartite systems, i.e. as in the usual way by simple tensor products.

\section{What Can be Measured at Accelerator Facilities?}\label{AcceleratorFacilities}

There are obviously two different questions that \textit{in principle} can be raised to the quantum system evolving in time:
\begin{itemize}
\item Are you a certain quasispin $|k_n\rangle$ at a certain time $t_n$ or not?
\item Or: Are you a certain quasispin $|k_n\rangle$ or its orthogonal state $|k_n^\perp\rangle$ ($\langle k_n^\perp|k_n\rangle=0$) at a certain time $t_n$?
\end{itemize}
where we denote by a quasispin $k_n$ any superposition of the mass eigenstates which are the solutions of the effective Schr\"odinger equation (\ref{schroedi}).

For non-decaying systems these questions are equivalent, but for decaying systems the second one means that you ignore all cases in which the neutral kaons decayed before the measurement, thus one does not take all information available into account. For studying certain quantum properties of these systems neglecting this kind of information is of no importance, however, e.g. if one is interested to show that there exists no explanation in terms of local hidden parameters for bipartite entangled decaying states, one is not allowed to selected only the surviving pairs, because one would not test the whole ensemble (consult Refs~\cite{Hiesmayr3,Hiesmayr13,Hiesmayr1} for more details).

Let us here also remark on what is meant by a measurement at a certain time $t_n$. Indeed, one does not measure time, but a certain final decay product or an interaction taking place at a certain position, point in space, in the detector. To be more precise one detects often only secondary reaction products and with the energy-momentum signature reconstructs the final states. Knowing the production point and thus the distance traveled as well as the momentum one can infer the proper time passed between production and decay or interaction.

There are in principle two different options which are denoted as an \textit{active} measurement procedure and a \textit{passive} measurement procedure, for reasons which may become clear in a moment, how to obtain the quasispin content of neutral mesons. This is a remarkable difference and gives raise to two further options of quantum erasure \cite{Hiesmayr7,Hiesmayr8} proving the very concept of a quantum eraser, i.e. sorting events to different available information. This kaonic quantum eraser is also in the future work programme of the upgraded KLOE detector which will start in 2011 (for a detailed program see Ref.~\cite{HiesmayrKLOE}).

For neutral kaons there exist two physical alternative bases. The first basis is the strangeness
eigenstate basis $\{| K^0\rangle, |\bar K^0 \rangle\}$. It can be measured by inserting along the
kaon trajectory a piece of ordinary matter. Due to strangeness conservation of the strong
interactions the incoming state is projected either onto $K^0$ by $K^0 p\rightarrow K^+ n$ or onto
$\bar K^0$ by $\bar K^0 p\rightarrow \Lambda \pi^+$, $\bar K^0 n\rightarrow \Lambda \pi^0$ or
$\bar K^0 n\rightarrow K^- p$. Here nucleonic matter plays the same role as a two channel analyzer
for polarized photon beams.

Alternatively, the strangeness content of neutral kaons can be determined by observing their
semileptonic decay modes (see Eq.(\ref{semileptonic-decays})). Obviously, the experimenter has no control over the kaon decay process, neither of the mode nor of the time.
The experimenter can only sort at the end of the day all observed events in proper decay modes and
time intervals. We call this procedure opposite to the \textit{active} measurement procedure
described above a \textit{passive} measurement procedure of strangeness.

The second basis $\{K_S,K_L\}$ consists of the short-- and long--lived states having well defined
masses $m_{S(L)}$ and decay widths $\Gamma_{(S)L}$, which are the solution of the Hamiltionian under investigation. It is the appropriate basis
to discuss the kaon propagation in free space, because these states preserve their own identity in
time. Due to the huge difference in the decay widths the short lived states $K_S$
decay much faster than the long lived states $K_L$. Thus in order to observe if a propagating kaon is a $K_S$ or
$K_L$ at an instant time $t$, one has to detect at which time it subsequently decays. Kaons which
are observed to decay before $\simeq t + 4.8\, \tau_S$ have to be identified as short lived states $K_S$, while
those surviving after this time are assumed to be long lived states $K_L$. Misidentifications reduce only to a few
parts in $10^{-3}$ (see also Refs.~\cite{Hiesmayr7,Hiesmayr8}). Note that the experimenter does not care
about the specific decay mode, she or he  records only a decay event at a certain time. This
procedure was denoted as an \textit{active} measurement of lifetime.

Neutral kaons are famous in Particle Physics as they violate the $\mathcal{CP}$ symmetry, where $\mathcal{C}$ stands for charge conjugation, i.e. interchanging a particle with an antipartice state and $\mathcal{P}$ for parity. So far no violation of the combined symmetry $\mathcal{CPT}$ has been found. Conservation of the $\mathcal{CPT}$ symmetry requires that the time reversal symmetry ${\cal T}$  has to be
broken. The break of the ${\cal T}$ invariance is far from being straightforwardly to be proven experimentally, because for a decay progress $A\longrightarrow B+C$ practical considerations prevent one from creating the time reversed sequence $B+C\longrightarrow A$. The CPLEAR collaboration was able to experimentally prove the ${\cal T}$ violation. At the first side it might be surprising that one finds a ${\cal T}$ violation in a framework which is completely controlled by non-relativistic quantum mechanics. The apparent paradox is resolved by remembering that the dynamics of a quantum system is given by the equation of motions and the boundary conditions. In particular, the fact that the relative weights of the mass eigenstates are different for the states of the two strangeness states leads to the observable effects. Or differently stated the ${\cal T}$ violation follows from the ${\cal CP}$ asymmetry in the initial states. Certainly, to understand and handle these symmetry violations we have to use the framework provided by relativistic quantum field theories. The author of Ref.~\cite{Vaccaro} argued that the measured ${\cal T}$ violation at accelerator facilities introduce destructive interference between different paths that the universe can take through time, she concludes that only two possible paths are surviving, one forward in time, the other one backward in time.

Since the neutral kaon system violates the ${\mathcal CP}$ symmetry (which will be discussed in Section \ref{CP})
the mass eigenstates are not strictly orthogonal, $\langle K_S|K_L\rangle\neq 0$. However,
neglecting ${\mathcal CP}$ violation ---it is of the order of $10^{-3}$--- the $K_S$'s are
identified by a $2\pi$ final state and $K_L$'s by a $3\pi$ final state. One denotes this procedure as a
\textit{passive} measurement of lifetime, since the kaon decay times and decay channels used in
the measurement are entirely determined by the quantum nature of kaons and cannot be in any way
influenced by the experimenter.

We have introduced two conceptually different procedures --\textit{active} and \textit{passive}-- to measure two different observables of the neutral kaon systems: strangeness or lifetime. The $\textit{active}$ measurement of strangeness is monitored by strangeness conservation in strong interactions while the corresponding \textit{passive} measurement is assured by the $\Delta S=\Delta Q$ rule, i.e. the change of the strangeness number and the change of the charge in a process. \textit{Active} and \textit{passive} lifetime measurements are efficient thanks to the smallness of $\frac{\Gamma_L}{\Gamma_S}$ and the ${\cal CP}$ violation parameter, respectively. This will be deeper analyzed in terms of the Heisenberg uncertainty relation in the entropic version in Section~\ref{Heisenberg}.

\textit{Active} measurements are possible due to a huge difference in lifetime of the two mesons and, therefore, in practice are available. Thus the neutral kaon system is special concerning its natural constance of the dynamic and, therefore, we mostly stick to this system.

The set of \textit{passive} measurements is not solely limited to the two above described basis choices, but are all possible decay modes of neutral mesons which e.g. single out different ${\cal CP}$ violation mechanisms. These decay modes can always be related to a certain quasispin at the moment of decay. Let us assume we find the final state $f$ at a time $t_n$ and we produced at time $t=0$ a quasispin $k_m$, the decay rate which is the derivative of the probability is given as an integral over the amplitude squared
\begin{eqnarray}
\Gamma(k_m(t_n)\longrightarrow f)=\int dph(f) |\langle f |{\texttt T}|k_m(t)\rangle|^2
\end{eqnarray}
where ${\texttt T}$ is the transition operator and the integral is taken over the phase space. To connect the quasipin with the final state, we have to require
\begin{eqnarray}
\langle k_n^\perp|k_n\rangle& \stackrel{!}{=}&0\qquad\textrm{and}\quad \langle f |{\texttt T}|k^\perp_n\rangle\stackrel{!}{=}0\;\longrightarrow P_{f}+P_{f^\perp}=1
\end{eqnarray}
and therefore any final decay product corresponds to a certain quasispin, i.e. a certain superposition of the mass eigenstates, e.g. a two pion event
corresponds to the quasispin
\begin{eqnarray}\label{k00}
|K_{\pi^0\pi^0}\rangle\equiv|k_n\rangle&=& \alpha_{00}\;|K_S\rangle+\beta_{00}\;|K_L\rangle\;.
\end{eqnarray}

Summarizing, we have for neutral kaons different conceptual measurement procedures if we neglect ${\cal CP}$ violation. \textit{Active} measurements are e.g. required when testing Bell inequalities (see Section \ref{BI}) while the existence of these two procedures opens new possibilities for kaonic quantum erasure experiments which have no analog for any other two-level quantum systems \cite{Hiesmayr7,Hiesmayr8} and are in the experimental programme of the KLOE-2 collaboration~\cite{HiesmayrKLOE}. If one is interested in other features of the quantum system under investigation or  including ${\cal CP}$ violation one can consider all decay channels. For example we will calculate the Heisenberg uncertainty due to ${\cal CP}$ violation in the case of two pion events (see Section \ref{Heisenberg}). If not stated differently we neglect $\mathcal{CP}$ violation.

\section{Effective Operators - A Heisenberg Picture for Decaying Systems}\label{effectiv}

To develop an effective formalism to derive any expectation value for the questions ``\textit{Are you in the quasispin $k_n$ at time $t_n$ (Yes) or not (No)}'' of decaying systems
\begin{eqnarray}
E(k_n,t_n)&=& P(\textrm{Yes}: k_n,t_n)-P(\textrm{No}: k_n,t_n)\nonumber\\
&\stackrel{P(\textrm{No}: k_n,t_n)+P(\textrm{Yes}: k_n,t_n)=1}{=}&2\; P(\textrm{Yes}: k_n,t_n)-1
\end{eqnarray}
we have to derive the probability to find a certain quasispin $k_n$ at time $t_n$ for a general initial state $\rho$, i.e.
\begin{eqnarray}
P(\textrm{Yes}: k_n,t_n)&=& Tr(\left(\begin{array}{cc}|k_n\rangle\langle k_n|&0\\
0&0\end{array}\right)\rho(t_n))\nonumber\\
&=&\rho_{SS}\cdot \cos^2\frac{\alpha_n}{2} e^{-\Gamma_S t_n}+\rho_{LL}\cdot \sin^2\frac{\alpha_n}{2} e^{-\Gamma_L t_n}\nonumber\\
&+&\rho_{SL}\cdot\cos\frac{\alpha_n}{2}\sin\frac{\alpha_n}{2}\;e^{i (\phi_n-t_n)}\cdot e^{-\Gamma t_n}\nonumber\\
&+&(\rho_{SL}\cdot\cos\frac{\alpha_n}{2}\sin\frac{\alpha_n}{2}\;e^{i (\phi_n-t_n)}\cdot e^{-\Gamma t_n})^*\;.
\end{eqnarray}
where we used the following parameterizations
\begin{eqnarray}
|k_n\rangle&=\cos\frac{\alpha_n}{2}|K_S\rangle+\sin\frac{\alpha_n}{2}\cdot e^{i\phi_n}|K_L\rangle\;.
\end{eqnarray}
and $\rho(t)$ is derived from the master equation (\ref{masterequation}). Moreover, we used a convenient re-scaling, i.e. $\Delta m:=1$ and, consequently the decay constants are re-scaled by the same factor $\Gamma_i:=\frac{\Gamma_i}{\Delta m}$.

From that we can extract a time dependent effective operator in dimensions $2\times 2$
\begin{eqnarray}
E(k_n,t_n)&=&Tr(O^{eff}(\alpha_n,\phi_n, t_n)\;\rho)
\end{eqnarray}
where $\rho$ is any initial state which can be taken in dimensions $2\times 2$ as at $t=0$ the decay products have not be taken into account. Herewith, we found for general decaying systems an effective operator in the Heisenberg picture which has
besides the computational and interpretative advantage a conceptual one (discussed in the following Sections), i.e. it generalizes for multipartite systems simply by the usual tensor product structure
\begin{eqnarray}
&&E(k_{n_1},t_{n_1};k_{n_1},t_{n_1};\dots;k_{n_k},t_{n_k})\\
&=& Tr(O^{eff}(\alpha_{n_1},\phi_{n_1}, t_{n_1})\otimes O^{eff}(\alpha_{n_2},\phi_{n_2}, t_{n_2})\otimes\dots\otimes O^{eff}(\alpha_{n_k},\phi_{n_k}, t_{n_k})\;\rho)\;.\nonumber
\end{eqnarray}
To derive these expectation values is rather cumbersome, e.g. for bipartite systems one has to derive the following four probabilities ($\texttt{P}_i=|k_i\rangle\langle k_i|$)
\begin{eqnarray}
P(Yes: k_n, t_n; Yes: k_m, t_m)&=&Tr_A(\texttt{P}_n \Lambda^{\textrm{single}}_{t_n}[Tr_B[\texttt{P}_m \Lambda^{\textrm{bipartite}}_{t_m}[\rho]]])\nonumber\\
P(Yes: k_n, t_n; No: k_m, t_m)&=&Tr_A(\texttt{P}_n \Lambda^{\textrm{single}}_{t_n}[Tr_B[(\mathbbm{1}-\texttt{P}_m) \Lambda^{\textrm{bipartite}}_{t_m}[\rho]]])\nonumber\\
P(No: k_n, t_n; Yes: k_m, t_m)&=&Tr_A((\mathbbm{1}-\texttt{P}_n) \Lambda^{\textrm{single}}_{t_n}[Tr_B[\texttt{P}_m \Lambda^{\textrm{bipartite}}_{t_m}[\rho]]])\nonumber\\
P(No: k_n, t_n; No: k_m, t_m)&=&Tr_A((\mathbbm{1}-\texttt{P}_n) \Lambda^{\textrm{single}}_{t_n}[Tr_B[(\mathbbm{1}-\texttt{P}_m) \Lambda^{\textrm{bipartite}}_{t_m}[\rho]]])\nonumber\\
\end{eqnarray}
to obtain the expectation value $E(k_n, t_n;k_m, tm)= P(Yes: k_n, t_n; Yes: k_m, t_m)+P(No: k_n, t_n; No: k_m, t_m)-P(Yes: k_n, t_n; No: k_m, t_m)-P(No: k_n, t_n; Yes: k_m, t_m)$,
where $\Lambda^{\textrm{single}}$ and $\Lambda^{\textrm{bipartite}}$ are the Liouville operators of the two master equations (\ref{masterequation}), respectively ($t_n>t_m$).

\subsection{What Observables are in Principle Accessible in Decaying Systems?}

Explicitly the effective operator for a two state decaying system decomposed into the Pauli matrices $\sigma$ is given by
\begin{eqnarray}
O^{eff}(\alpha_n,\phi_n, t_n)=-n_0(\alpha_n,t_n)\mathbbm{1}+\vec{n}(\alpha_n,\phi_n,t_n)\vec{\sigma}
\end{eqnarray}
with $\Delta\Gamma=\frac{\Gamma_L-\Gamma_S}{2}$
\begin{eqnarray}
\vec{n}(\alpha_n,\phi_n,t_n)= e^{-\Gamma t_n}\left(\begin{array}{c} \cos(t_n+\phi_n)\sin(\alpha_n)\\
\sin(t_n+\phi_n)\sin(\alpha_n)\\ \sinh(\Delta \Gamma t_n)+\cosh(\Delta\Gamma t_n)\cos\alpha_n\end{array}\right)
\end{eqnarray}
and $n_0(\alpha_n,t_n)=1-|\vec{n}(\alpha_n,\phi_n,t_n)|$. For spin $\frac{1}{2}$ systems, the most general observable is given by $\vec{n}\vec{\sigma}$ where any normalized quantization direction ($|\vec{n}|=1$) parameterized by polar angles $\alpha_n$ and $\phi_n$ can be chosen. In case of decaying systems we can choose in principle $\alpha_n$ and $\phi_n$ but for $t_n>0$ the ``quantization direction'' is no longer normalized and its loss results in an additional contribution in form of ``white noise'', i.e. the expectation value has a contribution independent of the initial state. \begin{eqnarray}
E(\alpha_n,\phi_n,t_n)&=&Tr(O^{eff}(\alpha_n,\phi_n, t_n)\rho)\nonumber\\
&=&-n_0(\alpha_n,t_n)+Tr(\vec{n}(\alpha_n,\phi_n,t_n)\vec{\sigma}\rho)\;.
\end{eqnarray}

One recognizes the involved role of the time-evolution: It is damping the ``Bloch'' vector $\vec{n}$ by $e^{-\Gamma t_n}$ and is responsible for the rotation or oscillation in the system, represented by the polar angle $\Phi=t_n+\phi_n$ in the $x,y$ equatorial plane ($x$ and $y$ component of the ``Bloch'' vector $\vec{n}$ corresponding to the strangeness eigenstates). In case of $\Delta\Gamma\not=0$ the time dependence of the $z$ component is more involved. This complex behaviour is responsible for certain quantum features of the system which we will analyze in the following part of the paper.

Let us discuss the eigenstates of the effective operator in order to gain a more physical intuition. For that we derive its spectral decomposition
\begin{eqnarray}\label{eigenvalueswcp}
O^{eff}_n &\equiv& O^{eff}(\alpha_n,\phi_n, t_n)\\
&=&(2 |\vec{n}(\alpha_n,\phi_n,t_n)|-1)\cdot|\chi(\alpha_n,\phi_n,t_n)\rangle\langle\chi(\alpha_n,\phi_n,t_n)|\nonumber\\
&&+(-1)\cdot|\chi(\alpha_n+\pi,\phi_n+2 t_n,-t_n)\rangle\langle\chi(\alpha_n+\pi,\phi_n+2 t_n,-t_n)|\nonumber
\end{eqnarray}
with
\begin{eqnarray}
|\chi_n\rangle&\equiv& |\chi(\alpha_n,\phi_n,t_n)\rangle\\
&=&\frac{1}{\sqrt{N(\alpha_n,t_n)}} \left\lbrace \cos\frac{\alpha_n}{2}\cdot \; e^{-\frac{\Gamma_S}{2} t_n}|K_S\rangle+ \sin\frac{\alpha_n}{2} \;e^{i(t_n+\phi_n)}\cdot e^{-\frac{\Gamma_L}{2} t_n}\;|K_L\rangle\right\rbrace
\nonumber\\
&&\textrm{with}\qquad N(\alpha_n,t_n)=
|\vec{n}(\alpha_n,\phi_n,t_n)|^2\;.\nonumber
\end{eqnarray}

The first eigenvector can be interpreted as a quasispin $k_n$ evolving in time according to the dynamics given by the non-hermitian Hamiltonian and normalized to surviving kaons, i.e. to \begin{eqnarray}
|\chi_n\rangle&\equiv&|k_n(t_n)\rangle\nonumber\\
&=&\frac{1}{\sqrt{N(\alpha_n,t_n)}}\{\cos\frac{\alpha_n}{2}\;e^{i \lambda_S^* t_n}|K_S\rangle+\sin\frac{\alpha_n}{2}\; e^{i\phi_n}\cdot e^{i \lambda_L^* t_n}\;|K_L\rangle\}\;.\end{eqnarray}
The second eigenvector related to the time--independent eigenvalue can be interpreted besides being orthogonal to the normalized quasispin $k_n$ as a quasispin evolving backward in time, but with no phase changes, which we discuss in the next Section \ref{CP} in more detail.

\subsection{{\cal CP} Violation in Mixing and the Effect on the Time Evolution}\label{CP}

In 1964 Cronin and Fitch discovered in a seminal experiment that in the neutral kaon system the symmetry ${\cal CP}$, where ${\cal C}$ stands for
charge conjugation, i.e. interchanging a particle state by an antiparticle state, and ${\cal P}$ is parity operator, is broken, for which they got the Nobel Prize in 1980. The ${\cal CP}$ violation (for a review see e.g. Ref.~\cite{Bigi2}) and its origin is still a hot discussed subject in particle physics. These open questions are addressed by recently approved projects as KLOE-2 and NA-62 for kaons and SuperBelle and SuperB for B--mesons.

The ${\cal CP}$ violation in mixing is e.g. measured by the semileptonic decay channels, i.e a strange quark $s$ decays weakly as
constituent of $\bar K^0\,$:

\vspace{0.1cm}

\begin{center}
\includegraphics[height=1cm]{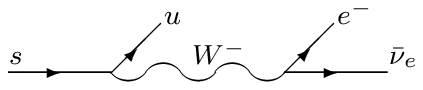}
\end{center}

\vspace{0.3cm}
Due to their quark content the kaon $K^0(\bar s d)$ and the anti--kaon $\bar K^0(s \bar d)$ have the following definite decay channels:
\begin{eqnarray}\label{semileptonic-decays}
K^0(d\bar{s}) \;&\longrightarrow&\; \pi^-(d\bar{u})\;\; l^+\;\nu_l \qquad
\textrm{where} \qquad \bar{s} \;\longrightarrow\; \bar{u}\;\; l^+\;\nu_l \nonumber \\
\bar{K}^0(\bar{d}s) \;&\longrightarrow&\; \pi^+(\bar{d}u)\;\; l^-\;\bar{\nu}_l \qquad
\textrm{where} \qquad  s \;\longrightarrow\; u\;\; l^-\;\bar{\nu}_l \;,
\end{eqnarray}
with $l$ either muon or electron, $l=\mu, e\,$.  Here the validity of the $\Delta S=\Delta Q$ rule is assumed. The Standard Model predicts negligible violations of this selection rule. When studying the leptonic charge asymmetry
\begin{eqnarray}\label{asymlept}
\delta &=& \frac{\Gamma(K_L\rightarrow \pi^- l^+ \nu_l) - \Gamma(K_L\rightarrow \pi^+ l^- \bar
\nu_l)}{\Gamma(K_L\rightarrow \pi^- l^+ \nu_l) + \Gamma(K_L\rightarrow \pi^+ l^- \bar \nu_l)} \;,
\end{eqnarray}
we notice that $l^+$ and $l^-$ tag $K^0$ and $\bar K^0$, respectively, in the $K_L$ state, and the
leptonic asymmetry (\ref{asymlept}) is expressed by the probabilities $|p|^2$ and $|q|^2$ of
finding a $K^0$ and a $\bar K^0$, respectively, in the $K_L$ state
\begin{eqnarray}
\delta &=& \frac{|p|^2-|q|^2}{|p|^2+|q|^2} \;,
\end{eqnarray}
i.e. the mass eigenstates and strangeness eigenstates are connected by
\begin{eqnarray}\label{kaonSL}
|K_S\rangle = \frac{1}{N}\big\lbrace p |K^0\rangle-q |\bar K^0\rangle \big\rbrace \;, \qquad
|K_L\rangle = \frac{1}{N}\big\lbrace p |K^0\rangle+q |\bar K^0\rangle \big\rbrace \;.
\end{eqnarray}
The weights $p=1+\varepsilon$, $\,q=1-\varepsilon$ with $N^2=|p|^2+|q|^2$ contain the complex
$CP$ \textit{violating parameter} $\varepsilon$ with $\lvert\varepsilon\rvert\approx10^{-3}$.
$CPT$ \textit{invariance} is assumed ($T \dots$ time reversal). The short--lived K--meson decays
dominantly into $K_S\longrightarrow 2 \pi$ with a width or lifetime $\Gamma^{-1}_S\sim\tau_S =
0.89 \times 10^{-10}$ s and the long--lived K--meson decays dominantly into $K_L\longrightarrow 3
\pi$ with $\Gamma^{-1}_L\sim\tau_L = 5.17 \times 10^{-8}$ s. However, due to ${\cal CP}$ violation we
observe a small amount $K_L\longrightarrow 2 \pi$. Therefore,
$CP$ violation expresses that there is a difference between a world of matter and a world of
antimatter.

Let us now derive the change due to $\mathcal{CP}$ violation to the effective observable. Firstly note that the length of the Bloch vector $\vec{n}$ can be
rewritten by the sum of two probabilities, i.e.
\begin{eqnarray}
|\vec{n}|&=&1-n_0\;=\;|\langle k_n|K_S(t_n)\rangle|^2+|\langle k_n|K_L(t_n)\rangle|^2\;.
\end{eqnarray}
The symmetry violation $\mathcal{CP}$ results in a non-orthogonality of the mass eigenstates, i.e. each amplitude leads to an interference term
\begin{eqnarray}
|\langle k_n|K_S(t_n)\rangle|^2&=&e^{\Gamma_S t_n}\;|\cos\frac{\alpha_n}{2}+\delta\cdot \sin\frac{\alpha_n}{2} e^{-i\phi_n}|^2\nonumber\\
|\langle k_n|K_L(t_n)\rangle|^2&=& e^{\Gamma_L t_n}\;|\delta\cdot\cos\frac{\alpha_n}{2}+ \sin\frac{\alpha_n}{2} e^{-i\phi_n}|^2
\end{eqnarray}
and, therefore, changes the oscillation behaviour of the system but as well the loss in the decaying system. Note that $\mathcal{CP}$ violation may as well change the state under investigation, i.e. the expectation value gets as well a ``contribution'' of the symmetry violation from the initial state.

The effective operator changes in detail by (we suppress the dependence on the parameters  $\alpha_n,\phi_n, t_n$)
\begin{eqnarray}
n^{\cal CP}_1&=&n_1-e^{-\Gamma t_n}(2\delta\cdot \cos t_n+\delta^2\cdot \sin\alpha_n\; \cos(t_n-\phi_n))\nonumber\\
n^{\cal CP}_2&=&n_2-e^{-\Gamma t_n}(2 \delta\cdot \sin t_n+\delta^2\cdot \sin\alpha_n\; \sin(t_n-\phi_n))\nonumber\\
n^{\cal CP}_3&=&n_3-(\delta\cdot (e^{-\Gamma_S t_n}-e^{-\Gamma_L t_n}) \sin\alpha_n\cos\phi_n\nonumber\\
&&\hphantom{n_3-(}+\delta^2\cdot \frac{1}{2}(e^{-\Gamma_S t_n}-e^{-\Gamma_L t_n}-(e^{-\Gamma_S t_n}+e^{-\Gamma_L t_n})\cos\alpha_n)\;.\nonumber\\
\end{eqnarray}
The spectral decomposition shows that the time dependent eigenvalue is changed by ${\cal CP}$ violation, confirming its observable character, but it has the same dependence from the Bloch vector as in case of ${\cal CP}$ conservation (compare with Eq.~(\ref{eigenvalueswcp}))
\begin{eqnarray}
\lambda_1^{\cal CP}&=&-1+2\; |\vec{n}^{\cal CP}|=1-2\; n_0^{\cal CP}\nonumber\\
\lambda_2^{\cal CP}&=&-1\;.
\end{eqnarray}
The two eigenvectors of the effective observable change accordingly
\begin{eqnarray}\label{cpeigenvectors}
|\chi_n^{{\cal CP},1}\rangle&=&
\frac{1}{\sqrt{N(t)}}
\biggl\lbrace \langle K_S|k_n\rangle\cdot e^{i\lambda_S^* t_n}\;|K_1\rangle+\langle K_L|k_n\rangle\cdot e^{i\lambda_L^* t_n}\;|K_2\rangle\biggr\rbrace\nonumber\\
|\chi_n^{{\cal CP},2}\rangle&=&
\frac{1}{\sqrt{N(-t)}}
\biggl\lbrace -\langle K_L|k_n\rangle^*\cdot e^{i\lambda_S t_n}\;|K_1\rangle+\langle K_S|k_n\rangle^*\cdot e^{i\lambda_L t_n}\;|K_2\rangle\biggr\rbrace\nonumber\\
\textrm{with}&&\nonumber\\
N(t)&=&e^{-\Gamma_S t_n} |\langle K_S|k_n\rangle|^2+e^{-\Gamma_L t_n} |\langle K_L|k_n\rangle|^2\;.
\end{eqnarray}
Note that if we parameterize the quasispin in the $\mathcal{CP}$ basis, $|k_n\rangle=\cos\frac{\alpha_n}{2}|K_1\rangle+\sin\frac{\alpha_n}{2}\cdot e^{i\phi_n t}\;|K_2\rangle$, we find that the weights do not add up to one generally
\begin{eqnarray}
N(0)&=&|\langle K_S|k_n\rangle|^2+|\langle K_L|k_n\rangle|^2\;=\; 1+\delta\cdot\sin\alpha_n\cos\phi_n\;.
\end{eqnarray}

\section{The Entropic Uncertainty Relation for Single and Bipartite Systems}\label{Heisenberg}

The entropic uncertainty relation of two non-degenerate observables is given by (introduced by D. Deutsch \cite{Deutsch}, improved in Ref.~\cite{Kraus} and proven by Ref.~\cite{MaassenUffink})
\begin{eqnarray}\label{EQUI}
H(O^{eff}_n)+H(O^{eff}_m)&\geq& 
- 2\log_2\left(\max_{i,j}\{|\langle\chi_n^i|\chi_m^j\rangle|\}\right)
\end{eqnarray}
where  \begin{eqnarray}
H(O^{eff}_n)=-p(n)\log_2 p(n)-(1-p(n))\log_2 (1-p(n))
\end{eqnarray}
is the binary entropy for a certain prepared pure state $\psi$ and the $p(n)$'s are the probability distribution associated with the measurement of $O^{eff}_n$ for $\psi$,  hence $p(n)=|\langle \chi_n|\psi\rangle|^2$. This is a reformulation of the famous uncertainty principle by Robertson \cite{Robertson}, which can be found in most textbooks on quantum theory
\begin{eqnarray}
(\Delta O^{eff}_n)_\psi\cdot(\Delta O^{eff}_m)_\psi &\geq&\frac{1}{2}\; \left|\langle\psi|\left[O^{eff}_n,O^{eff}_m\right]|\psi\rangle\right|\;,
\end{eqnarray}
where  $(\Delta A)^2_\psi=\langle A^2\rangle_\psi-\langle A\rangle^2_\psi$ are the mean square deviations. Choosing, the operators, position $\hat{x}$ and momentum $\hat{p}$, the Robertson relation turns into the famous Heisenberg relation
\begin{eqnarray}
\Delta\hat{x}\cdot\Delta\hat{p}\geq\frac{1}{2}\;.
\end{eqnarray}

The maximal value of the right hand side of the entropic uncertainty relation is obtained for
\begin{eqnarray}
|\langle \chi_n|\chi_m\rangle|&=&\frac{1}{\sqrt{2}}\;,
\end{eqnarray}
in this case the the two observables are commonly called complementary to each other (their eigenvalues have to be nondegenerate), e.g. if the operators are $\sigma_x$ and $\sigma_z$. In general a non-zero value of the right hand side of Eq.(\ref{EQUI}) means that the two observables do not commute, i.e. it quantifies the complementarity of the observables. The binary entropies on the left hand side quantify the gain of information on average when we learn about the value of the random variable associated to $O^{eff}_n$. Alternatively, one can interpret the entropy as the uncertainty \textit{before} we obtain the result of the random variable.

The reformulation of the Heisenberg relation (\ref{EQUI}) has ---besides its different information-theoretic interpretation and its stronger bound~\cite{Iwo}--- the advantage that the right hand side of the inequality is independent of the prepared state and only depends on the eigenvectors of the observables, hence puts a stronger limit on the extent to which the two observables can be simultaneously peaked.

Remarkably, the right hand side of the entropic uncertainty relation also does not depend on the eigenvalues (except to test the non-degeneracy), this means that if the state is prepared in an eigenstate say of $O^{eff}_n$ then the two eigenvalues of $O^{eff}_m$ are equally probable as measured values, i.e. the exact knowledge of the measured value of one observable implies maximal uncertainty of the measured value of the other, independent of the eigenvalues.

\subsection{An Information Theoretic View on Measurements at Different Times at Accelerator Facilities}

Particle detectors at accelerator facilities detect or reconstruct different decay products at different distances from the creation point, usually by a \textit{passive} measurement procedure, more rarely by an \textit{active} measurement procedure. Let us here discuss what is learnt by finding a certain quasispin $|k_n\rangle$ at a certain time $t_n$ or not which can correspond to a certain decay channel, compared to the situation to find a $k_m$ at the creation point $t_m=0$ or not. Certainly, this result also quantifies our uncertainty \textit{before} we learn the result (Yes, No) at $t_n$ and (Yes, No) at $t_m$. In particular, if we compare observables of same quasispin at different time, we obtain the uncertainty due to the time evolution.

Differently stated, we can view it in the following way~\cite{Renato}, two experimenters, Alice and Bob, choose two different measurements corresponding to the observables $O^{eff}_n, O^{eff}_m$. Alice prepares a certain state $\psi$ and sends it to Bob. Bob carries out one of the two measurements  $O^{eff}_n, O^{eff}_m$ and announces his choice $n$ or $m$ to Alice. She wants to minimize her uncertainty about Bob's measurement result. Alice's result is bounded by the equation~(\ref{EQUI}).

In case of unstable systems the right hand side of the entropic uncertainty relation (\ref{EQUI}), for which we have to find the maximum, is given by
\begin{eqnarray}
&&\max\biggl\{\langle\chi_m^1|\chi_n^1\rangle,\langle\chi_m^1|\chi_n^2\rangle,\langle\chi_m^2|\chi_n^1\rangle,\langle\chi_m^2|\chi_n^2\rangle\biggr\}
\end{eqnarray}
with $|\chi_n^1\rangle=|\chi(\alpha_n,\phi_n,t_n)\rangle$ and $|\chi_n^2\rangle=|\chi(\alpha_n+\pi,\phi_n+2 t_n,-t_n)\rangle$ being the eigenvectors of the effective operators or the quasispin propagating forward or backward in time, respectively. Any product derives to
\begin{eqnarray}
&&\langle \chi(\alpha_n,\phi_n,t_n)|\chi(\alpha_m,\phi_m,t_m)\rangle=\nonumber\\
&&\frac{\cos\frac{\alpha_n}{2}\cos\frac{\alpha_m}{2}+\sin\frac{\alpha_n}{2}\sin\frac{\alpha_m}{2}\; e^{i(t_m-t_n+\phi_m-\phi_n)}\cdot e^{-\Delta\Gamma(t_n+t_m)}}{\frac{1}{2}\sqrt{1+e^{-2\Delta \Gamma t_n}+\cos\alpha_n(1-e^{-2\Delta\Gamma t_n})}\sqrt{1+e^{2\Delta \Gamma t_m}-\cos\alpha_m(1-e^{2\Delta\Gamma t_m})}}\;.\nonumber\\
\end{eqnarray}

\begin{figure}
\centering
(a)\includegraphics[scale=0.5]{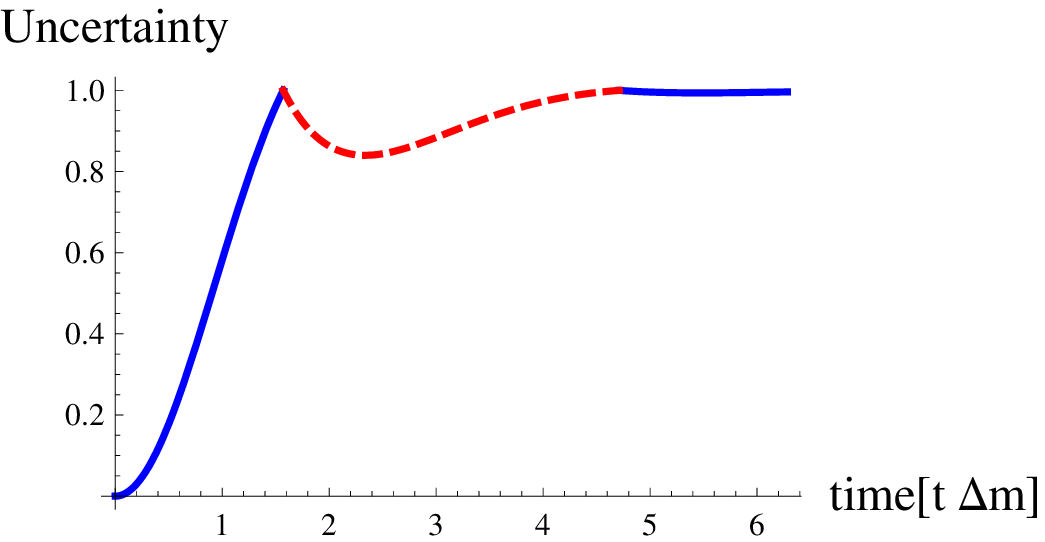}
(b)\includegraphics[scale=0.5]{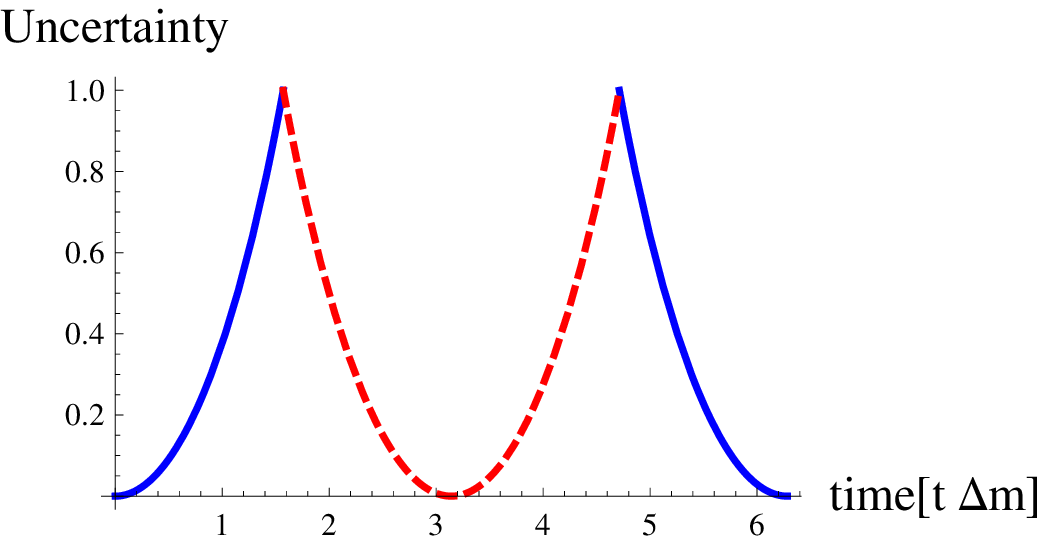}
\caption{Here the lower bound of the entropic quantum uncertainty inequality (\ref{EQUI}) is plotted in case of a strangeness event at $t=0$ compared to a strangeness event at a later time, i.e. for the observables $A=O_{eff}(\frac{\pi}{2},\phi_n,0)$ and $B=O_{eff}(\frac{\pi}{2},\phi_m,t)$ with $\phi_n=\phi_m=0,\pi$ for (a) $\Gamma_S$  and (b) $\Gamma_\approx\Gamma_L$ including $\Gamma_S=\Gamma_L=0$. The solid blue line shows when the eigenvectors both propagating forward in time or both propagating backward in time overlap maximally, whereas the red dashed line shows the case when forward and backward propagating quasipins overlap. Figure (b) shows the case of a slow decaying system or all other meson systems, i.e. $B_d, B_s$, except maybe the $D$ meson system for which not much precise data is available. If the decay constants are considerably different there is always missing information in the system.
}\label{uncertainty1}
\end{figure}

In Fig.~\ref{uncertainty1} we plotted the complementarity for the observable asking the question ``\textit{Is the neutral kaon system in the state $|K^0\rangle$ or not at time $t=0$}'' compared to the question ``\textit{Is the neutral kaon system in the state $|K^0\rangle$ or not at time $t$}'', i.e. comparing the complementary introduced by the time evolution in the case of strangeness measurements. Here Fig.~\ref{uncertainty1}~(a) refers to the neutral kaon case and $(b)$ to a slowly decaying system ($\Gamma_S\rightarrow 100 \Gamma_S$) or any of the other meson systems $\Delta\Gamma=0$. One notices that for times being odd multiples of $\frac{\pi}{2}$ the complementary of the two observables becomes minimal, while for even multiples it maximizes.

\begin{figure}
\centering
(a)\includegraphics[scale=0.5]{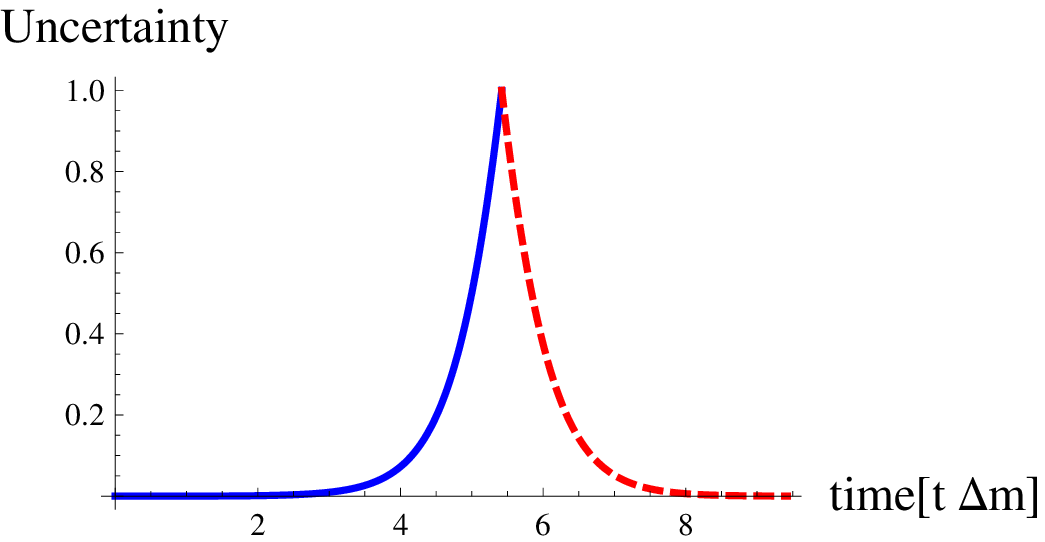}
(b)\includegraphics[scale=0.5]{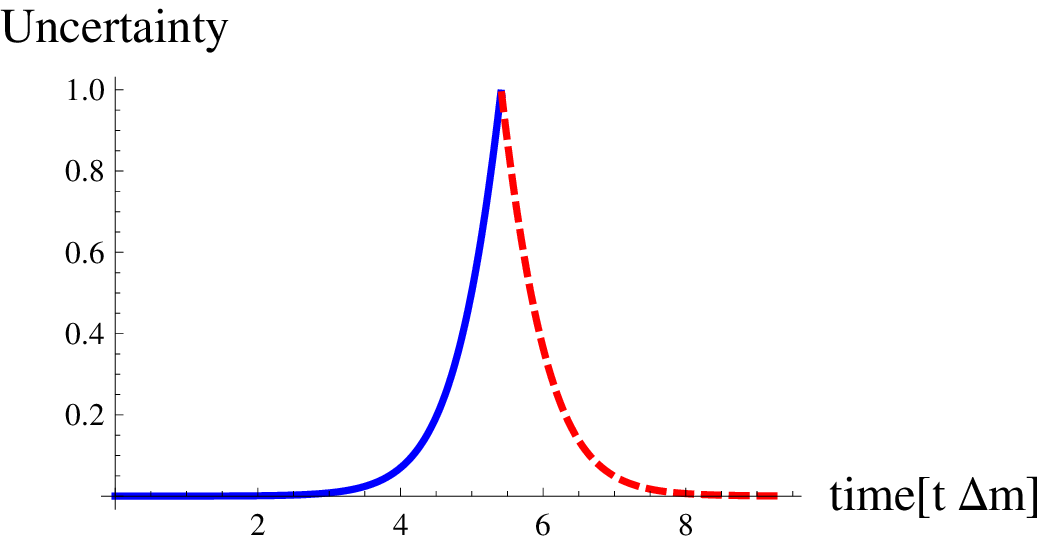}
(c)\includegraphics[scale=0.5]{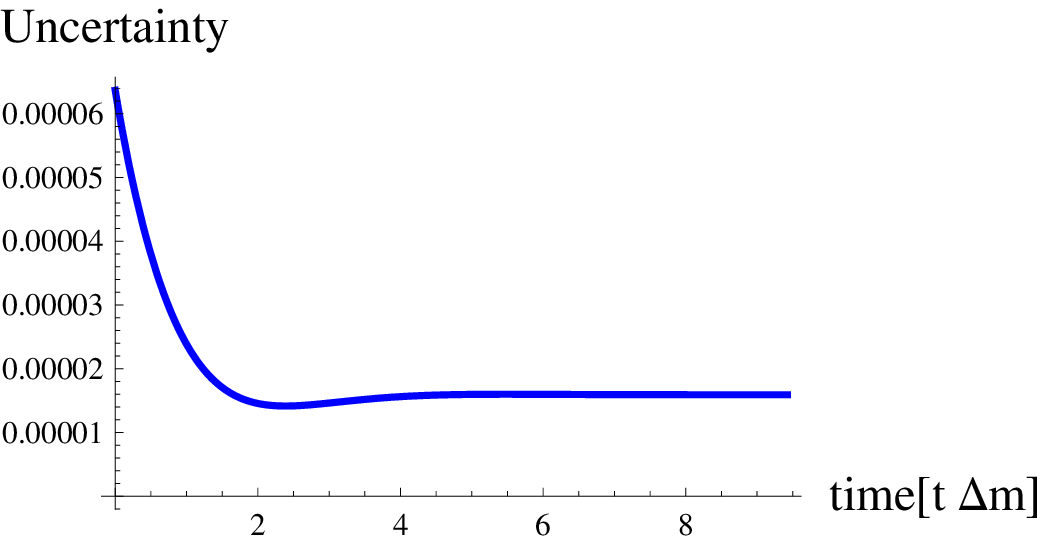}
(d)\includegraphics[scale=0.5]{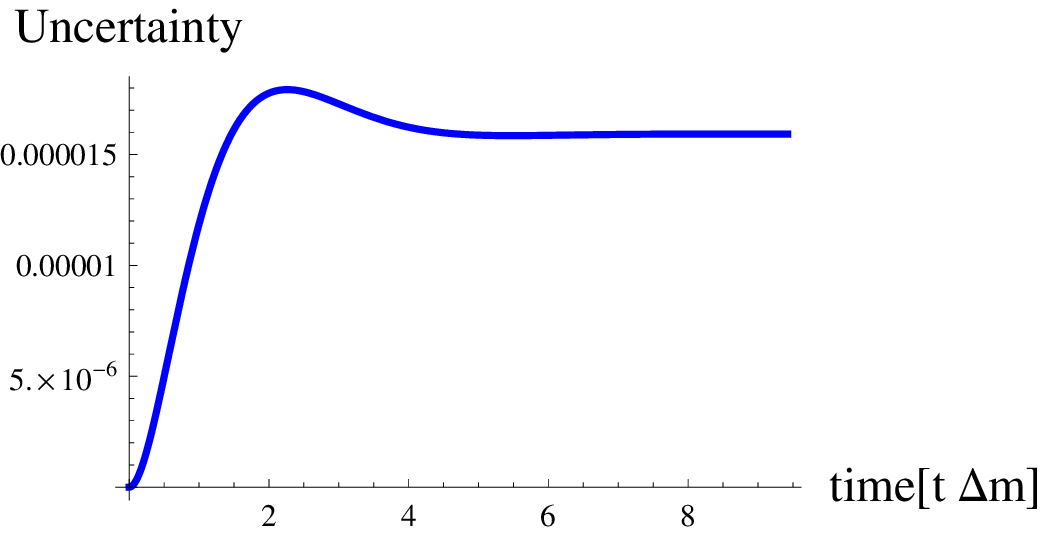}
\caption{This graphs depict the lower bound of the entropic quantum uncertainty inequality (\ref{EQUI}) by comparing measurements at time $t=0$ to measurements at later times $t$ for (a) short lived state ($t=0$) versus short lived state at $t$, (b) long lived state ($t=0$) versus short lived state at $t$, (c) short lived state ($t=0$) versus long lived state at $t$ and (d) long lived state ($t=0$) versus long lived state at $t$. This shows the uncertainty introduced by breaking the $\mathcal{CP}$ symmetry in the time evolution. If Alice and Bob agree to ask about a short lived state at the complementary time $t=5.4 [t \Delta m]\equiv 11.4 \tau_S$ the uncertainty becomes the maximal possible value. In case Alice and Bob agree to ask for any time $t>0$ for a long lived state, the uncertainty is nonzero.}
\label{uncertainty2}
\end{figure}

 Asking about the mass-eigenstates we find no complementary of the observables for any time, this certainly only changes if we include $CP$ violation. The uncertainty, i.e. the overlap of the measurement of a short lived state at a later time point to that at time zero, is moderated by $\delta$, i.e. for small times the maximum is obtained by the overlap of the first two eigenvalues Eq.(\ref{cpeigenvectors})
 \begin{eqnarray}
\left|\langle \chi^{{\cal CP},1}(K_S, t_n)|\chi^{{\cal CP},1} (K_S,t_m=0)\rangle\right| &=&\frac{\left|e^{-\frac{\Gamma_S}{2} t_n}+\delta^2 e^{-i t_n}\cdot e^{-\frac{\Gamma_L}{2} t_n}\right|}{\sqrt{(1+\delta^2)(e^{-\Gamma_S t_n} +\delta^2 e^{-\Gamma_L t_n})}}
 \end{eqnarray}
 and the maximum uncertainty $- 2\log_2\max\{|\langle\chi_n^i|\chi_m^j\rangle|\}$ is reached for the overlap $\frac{1}{\sqrt{2}}$ for $t_n=11.4 \tau_S$ and choosing the $\mathcal{CP}$ violation parameter $\delta=3.322\cdot 10^{-3}$ (world average \cite{PDG}). This is just the case when the overlaps of all possibilities are equal, i.e. the two bases are mutually unbiased bases $(MUBS)$.  The same complementary time $t_n=11.4 \tau_S$ is obtained when we compare the measurement of the long lived state at time $t_m=0$ and a measurement of the short lived state. For the two other options the maximal uncertainty can never be reached. Initially, the uncertainty is zero in case of measuring long lived states and then oscillates due to $\delta$ and reaches after $t_n=11.4 \tau_S$ a constant value close to zero. This is summarized in Fig.~\ref{uncertainty2}.

 Certainly, at this time the probability to find a short lived state is for all practical purposes zero. Remember that we have chosen for active measurements of lifetime a time of $4.8 \tau_S$, which is the time when the probability of not finding a short lived state when it was produced as a short lived state equals the probability to find a long lived state when it was produced as a long lived state, i.e. $1-e^{-\Gamma_S t}\stackrel{!}{=}e^{-\Gamma_L t}$. This time is by more than a factor $2$ different to the complementary time which strongly depends on the amount of $\mathcal{CP}$ violation.   We can revert the issue and ask how big $\delta$ needs to be in order that the two times would be equal: it would need to be $25$ times the value of $\delta$. Therefore, \textit{active} and \textit{passive} measurements of lifetime are efficient.

\subsection{The Uncertainty of Measurement Settings for Bipartite Kaons}

The effective operator formalism guarantees that the tensor product structure is conserved, i.e. the most general expectation value of a bipartite system is given by
\begin{eqnarray}
E(k_n,k_m)=Tr(O_n^{eff}\otimes O_m^{eff}\rho)
\end{eqnarray}
for any initial bipartite state $\rho$. In this case one studies e.g. symmetry violations or Bell inequalities where one compares measurements of different quasispins at different times. In this section we want to investigate the uncertainty of such different measurement settings and herewith obtain a different view and intuition on how certain properties of quantum states are revealed, in particular
we will then proceed to analyze the maximum violation of a Bell inequality.

To compute the right hand side of the entropic uncertainty relation we have to find the maximum of all eigenvectors of the operator $O_n\otimes O_m$ which is straightforward as it is simply the product of the eigenvectors of the single operators $O_{n/m}$ of Alice and Bob, respectively
\begin{eqnarray}
&&\max\biggl\{\langle\chi_m^i|\chi_n^j\rangle\cdot\langle\chi_m^k|\chi_n^l\rangle\biggr\}\quad\textrm{with}\;i,j,k,l=1,2\;.
\end{eqnarray}

\begin{figure}
\centering
(a)\includegraphics[scale=0.5]{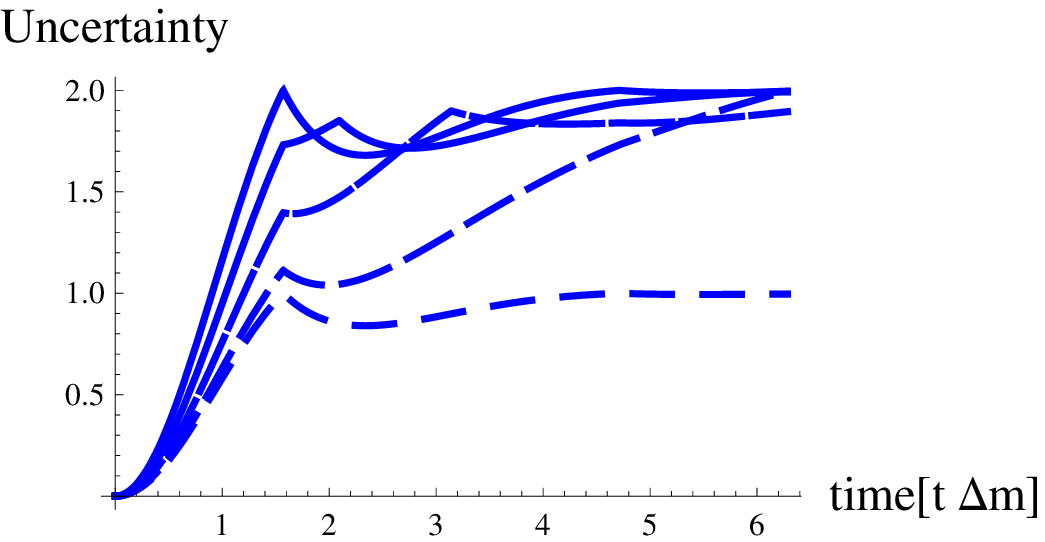}
(b)\includegraphics[scale=0.5]{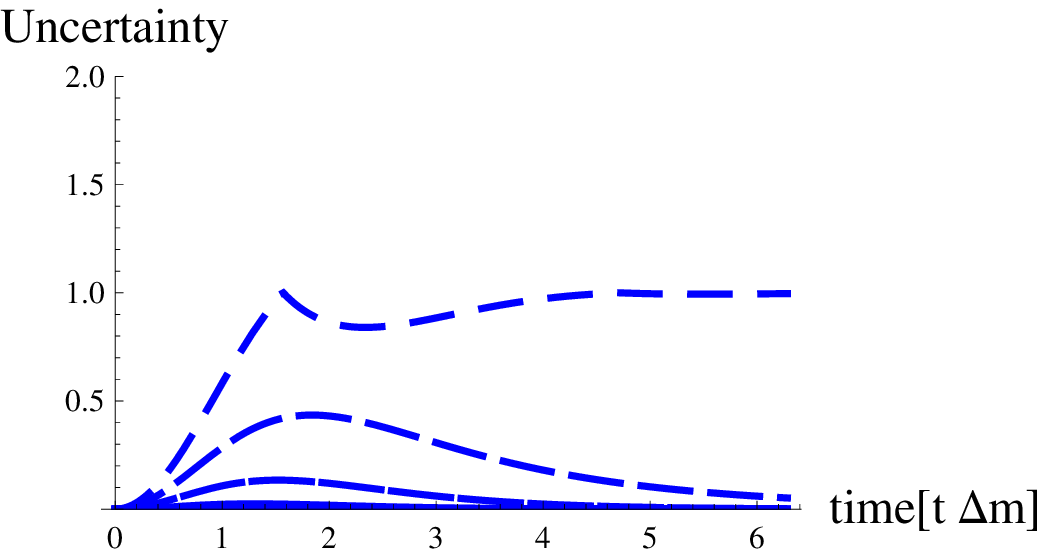}
\caption{The right hand side of the entropic quantum uncertainty inequality (\ref{EQUI})
for the two observables (a)
$O_{eff}(\frac{\pi}{2},0,0)\otimes O_{eff}(\frac{\pi}{2},0,t)$ versus $O_{eff}(\frac{\pi}{2},0,t_1)\otimes O_{eff}(\frac{\pi}{2},0,0)$ and (b)
 $O_{eff}(\frac{\pi}{2},0,0)\otimes O_{eff}(\frac{\pi}{2},0,t)$ versus $O_{eff}(\frac{\pi}{2},0,0)\otimes O_{eff}(\frac{\pi}{2},0,t_1)$ for $t_1=0,t/4,\dots,t$
 is plotted, where $t_1=0$ is the dashed line. One recognizes that one can increase or decrease the maximal uncertainty if the role of the first and second observable
 in the tensor product, i.e. Alice and Bob's role, is changed.
}\label{uncertainty4}
\end{figure}

In Fig.~\ref{uncertainty4} we show how the uncertainty is changed for different observables in the bipartite kaons system, which gives an intuition when a certain Bell operator may yield a violation (see next Section \ref{BI}).

\section{The Bell-CHSH Inequality}\label{BI}

In accelerator experiments one can produce a spin
singlet state, e.g. by the decay of a $\Phi$ meson at the DAPHNE machine. One has the same scenario as Einstein, Podolsky
and Rosen considered in 1935 which we write down for
different quantum systems (spin--$\frac{1}{2}$, ground/excited
state, polarisation, K--meson, B--mesons, molecules arriving early/late \cite{Hornberger} or single neutrons in an
interferometer) to show its similarity:
\begin{eqnarray}\label{antisymmetricBellstate}
|\psi^-\rangle&=&\frac{1}{\sqrt{2}}\big\lbrace
|\Uparrow\rangle_l\otimes|\Downarrow\rangle_r-|\Downarrow\rangle_l\otimes|\Uparrow\rangle_r\big\rbrace\nonumber\\
&=&\frac{1}{\sqrt{2}}\big\lbrace
|0\rangle_l\otimes|1\rangle_r-|1\rangle_l\otimes|0\rangle_r\big\rbrace\nonumber\\
&=&\frac{1}{\sqrt{2}}\big\lbrace
|H\rangle_l\otimes|V\rangle_r-|V\rangle_l\otimes|H\rangle_r\big\rbrace\nonumber\\
&=&\frac{1}{\sqrt{2}}\big\lbrace
|K^0\rangle_l\otimes|\bar K^0\rangle_r-|\bar K^0\rangle_l\otimes|K^0\rangle_r\big\rbrace\nonumber\\
&=&\frac{1}{\sqrt{2}}\big\lbrace
|B^0\rangle_l\otimes|\bar B^0\rangle_r-|\bar B^0\rangle_l\otimes|B^0\rangle_r\big\rbrace\nonumber\\
&=&\frac{1}{\sqrt{2}}\big\lbrace
|late\rangle_l\otimes| early\rangle_r-| early\rangle_l\otimes|late\rangle_r\big\rbrace\nonumber\\
&=&\frac{1}{\sqrt{2}}\big\lbrace
|I\rangle_l\otimes|\Uparrow\rangle_r-|II\rangle_l\otimes|\Downarrow\rangle_r\big\rbrace\nonumber\\
&=&\dots\;.
\end{eqnarray}
Analog to entangled photon systems for these systems Bell
inequalities can be derived, i.e. the most general Bell inequality
of the CHSH--type is given by (see Ref.~\cite{Hiesmayr4})
\begin{eqnarray}\label{chsh}
\lefteqn{S_{k_n,k_m,k_{n'},k_{m'}}(t_1,t_2,t_3,t_4)=}\nonumber\\
&&\left|
E_{k_n,k_m}(t_1,t_2)-E_{k_n,k_{m'}}(t_1,t_3)\right|\nonumber\\
&&\qquad\quad+|E_{k_{n'},k_{m}}(t_4,t_2)+E_{k_{n'},k_{m'}}(t_4,t_3)|\leq
2\;.
\end{eqnarray}
Here Alice can choose on the kaon propagating to her left hand side
to raise the question if the neutral kaon is in the quasispin $|k_n\rangle=\cos\frac{\alpha_n}{2} |K^0\rangle+\sin\frac{\alpha_n}{2} e^{i\phi_n} |\bar K^0\rangle$ or not,
and how long the kaon propagates, the time $t_n$. The same options are
given to Bob for the kaon propagating to the right hand side. As in
the usual photon setup, Alice and Bob can choose among two settings.

Differently to commonly investigated systems one has more options. One can vary in the quasispin space or
vary the detection times or both.

With our effective framework we can rewrite the Bell-CHSH-inequality in a witness type, i.e. with the Bell operator
\begin{eqnarray}\label{Belloperator}
\textbf{Bell}^{eff}&=& O^{eff}_n\otimes (O^{eff}_m-O^{eff}_{m'})+O^{eff}_{n'}\otimes (O^{eff}_m+O^{eff}_{m'})
\end{eqnarray}
any local realistic hidden parameter theory has to satisfy
\begin{eqnarray}
|Tr(\textbf{Bell}^{eff}\rho)|&\leq&2\;.
\end{eqnarray}
This operator form of the generalized Bell-CHSH inequality \cite{Hiesmayr3} gives us the opportunity to find for a given choice of Bell settings without optimization over all possible initial states whether the Bell inequality can be violated. In particular, the eigenvalues of the Bell operator give us the upper and lower bound that can be reached for the optimal initial state, i.e. the one which maximizes or minimizes the Bell inequality. Determining
whether a Bell inequality is preserved or violated for a given state $\rho$ is in general a high-dimensional
nonlinear constrained optimization problem. In Ref.~\cite{Spengler} a numerical method was shown by introducing a proper parameterization \cite{Spengler2} for unitary matrices to derive bounds on Bell inequalities for any qudit system ($d$--level system). This certainly is a benefit of our introduced effective formalism as optimization in this case is not needed. In any numerical optimization there is no guarantee that the global extremum was reached. In some exemplary cases we checked for the agreement and in many case the optimization failed.

We present first a generalized Bell inequality which has been discussed in literature \cite{Hiesmayr3,Hiesmayr4,Hiesmayr2,Hiesmayr1} and shows a relation between $\mathcal{CP}$ violation and the nonlocality detected by the above Bell inequality. Then we proceed to a Bell setting that can be realized in a direct experiment.

\subsection{A Bell Inequality Sensitive to  $\mathcal{CP}$ Violation}

Let us choose all times equal zero and choose the quasispin
states $k_n=K_S, k_m=\bar K^0, k_{n'}=k_{m'}=K_1^0$ where $K_1^0$ is the ${\cal CP}$ plus eigenstate.

In
Ref.~\cite{Hiesmayr4} the authors showed that after optimization the
CHSH--Bell inequality  can be turned for an initial spin singlet state into
\begin{eqnarray}\label{BICP}
\delta\leq 0\;
\end{eqnarray}
where $\delta$ is the ${\cal CP}$ violating parameter in mixing, Eq.(\ref{asymlept}).
Experimentally, $\delta$ corresponds to the leptonic asymmetry of
kaon decays which is measured to be $\delta=(3.322\pm0.055)\cdot
10^{-3}$. This value is in clear contradiction to the value required
by the CHSH-Bell inequality above, i.e. by the premises of local realistic theories!
The result can be also made stronger by changing the Bell setting by $K_S\longrightarrow K_L$, then one obtains $\delta\geq 0$, thus both CHSH-Bell inequalities require
\begin{eqnarray}
\delta=0\;,
\end{eqnarray}
i.e. any local realistic hidden variable theory is in contradiction to $\cal{CP}$ violation, a difference of a world of particles and antiparticles.
In this sense the violation of a symmetry in high energy physics is
connected to the violation of a Bell inequality, i.e. to
nonlocality. This is clearly not available for photons, they do not
violate the $\cal{CP}$ symmetry.

We also want to remark that the considered Bell inequality, since it is chosen at time $t=0$
is connected to a test of contextuality rather than nonlocality. \textit{Noncontextuality}, the
independence of the value of an observable on the experimental context due to its predetermination
---a main hypothesis in hidden variable theories--- is definitely ruled out! So the contextual
quantum feature is demonstrated for entangled kaonic qubits.

Although the Bell inequality~(\ref{BICP}) is as loophole free as possible, the
probabilities or expectation values involved are not directly
measurable, because experimentally there is no way to distinguish
the short--lived state $K_S$ from the ${\cal CP}$ plus state $K_1^0$
directly.

\subsection{A Bell Inequality Sensitive to Strangeness}

Let us now proceed to another choice for the Bell
inequality (\ref{chsh}), i.e. all quasispins equal $\bar K^0$, but we
are going to vary all four times
\begin{eqnarray}\label{BIstrangeness}
\lefteqn{S_{\bar K^0,\bar K^0,\bar K^0,\bar
K^0}(t_1,t_2,t_3,t_4)=}\nonumber\\
&&| E(\bar K^0, t_1; \bar K^0, t_2)-E(\bar K^0, t_1;\bar K^0, t_3)|+|E(\bar K^0, t_4; \bar K^0, t_2)+E(\bar K^0, t_4; \bar K^0, t_3)|\nonumber\\
&&\qquad\qquad\leq 2\;.
\end{eqnarray}
This has the advantage that it can in principle be tested in
experiments. Alice and Bob insert at a certain distance from the
source (corresponding to the detection times) a piece of matter
forcing the incoming neutral kaon to react. Because the strong
interaction is strangeness conserving one knows from the reaction
products if it is an antikaon or not. Note that different to photons
a $NO$ event does not mean that the incoming kaon is a $K^0$ but
also includes that it could have decayed before. In principle the
strangeness content can also be obtained via decay modes, but Alice
and Bob have no way to force their kaon to decay at a certain time,
the decay mechanism is a spontaneous event. However, a necessary
condition to refute local realistic theories are \textit{active}
measurements, i.e. exerting the free will of the experimenter (for
more details consult \cite{Hiesmayr13}).

In Refs.~\cite{Hiesmayr4,Hiesmayr13} the authors studied the problem
for an initial maximally entangled Bell state, i.e., $\psi^-\simeq
K^0 \bar K^0-\bar K^0 K^0$, and found that a value greater than $2$
cannot be reached, i.e. one cannot refute any local realistic
theory. The reason is that the particle--antiparticle oscillation is
too slow compared to the decay or vice versa, i.e., the ratio of
oscillation to decay $x=\frac{\Delta m}{\Gamma}$ is about $1$ for
kaons and not $2$ necessary for a violation. A different view is
that the decay property acts as a kind of ``decoherence'' as we introduced in Section \ref{timeevolution}. From
decoherence studies we know that some states are more ``robust''
against a certain kind of decoherence than others, this leads to the
question if another maximally entangled Bell state or maybe a
different initial state would lead to a violation which is indeed
the case.

In Ref.~\cite{Hiesmayr1} it was shown that such states exists. This shifts the problem to finding methods to produce these initial states leading to a violation of the generalized CHSH--Bell inequality. This is still an open problem. In Ref.~\cite{Hiesmayr1} also the interplay between entanglement and entropy was studied and as also shown by the authors of Ref.~\cite{Mazzola}, who studied the dynamics of two qubits interacting with a common zero-temperature non-Markovian reservoi, the picture that entanglement loss due to environmental decoherence is accompanied by loss of the purity of the state of the system does not apply to these systems.

\begin{figure}
(a)\includegraphics[scale=0.9]{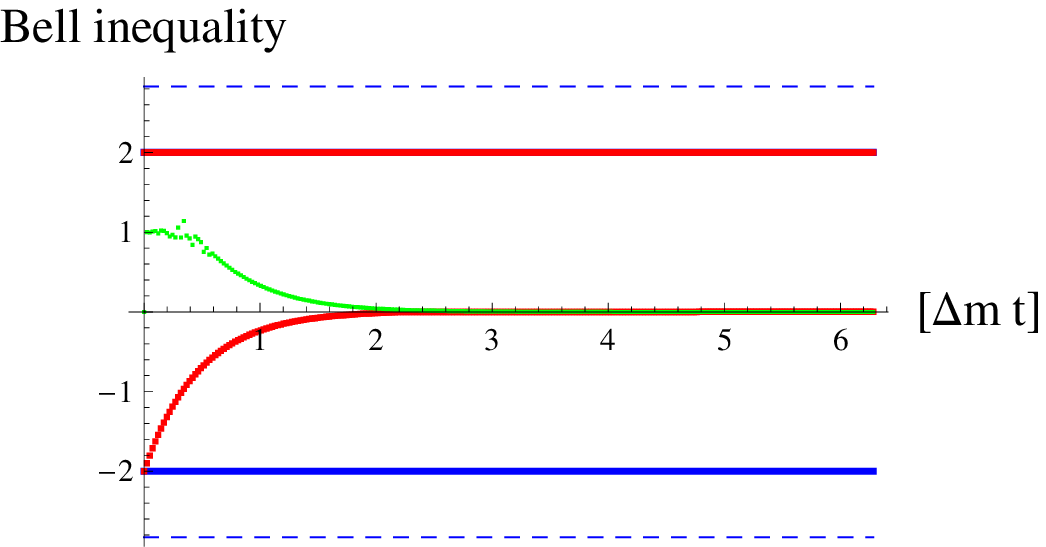}
(b)\includegraphics[scale=0.9]{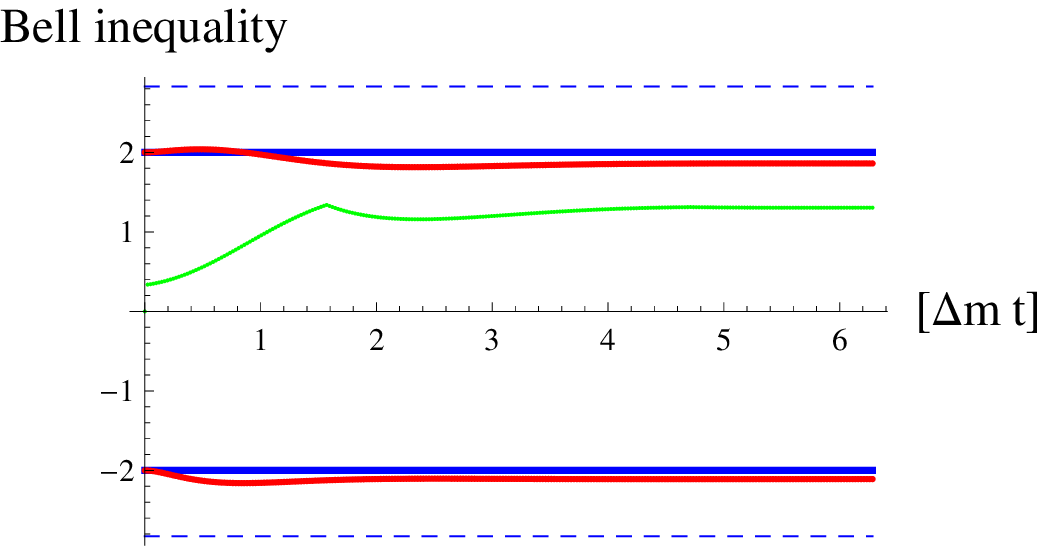}
(c)\includegraphics[scale=0.9]{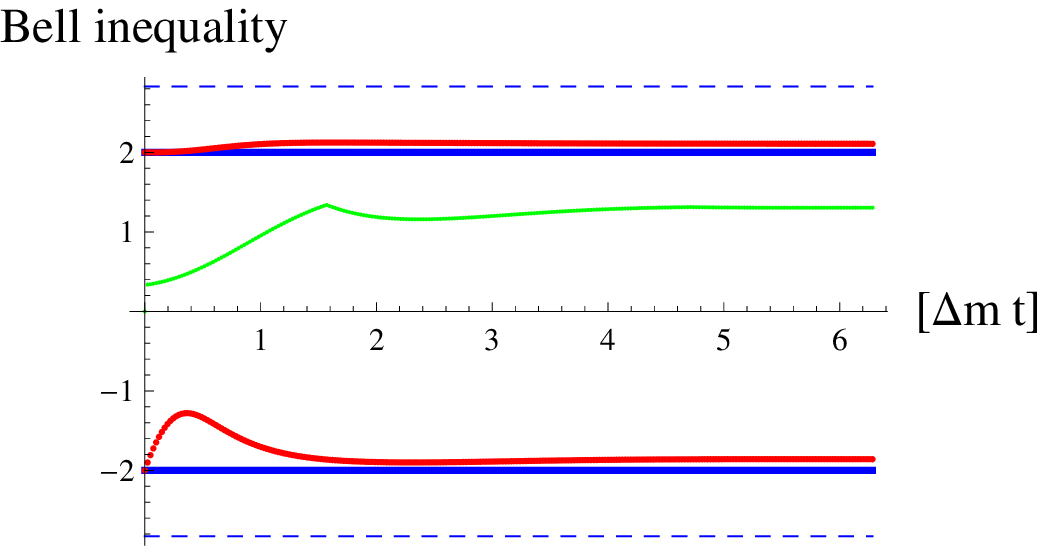}
\caption{(Color online) The maximal violations of the Bell inequality, i.e. the maximal and minimal eigenvalues of the operator (\ref{Belloperator}) for strangeness questions for time choices (a) $\{t_n=t_m=t_{n'}=t_{m'}=t\}$, (b) $\{t_n=0,t_m=t,t_{n'}=t,t_{m'}=0\}$ and (c) $\{t_n=t,t_m=0,t_{n'}=0,t_{m'}=t\}$ are plotted (red big dots). Green dots (smaller dots) represent a lower bound on the entropic uncertainty relation (\ref{EQUI}) between the two summands of the Bell operator which is zero for $t=0$ and then immediately jumps to a certain value and is equal for the time settings (b) and (c). The dashed blue lines are the upper bounds on the CHSH-Bell inequality, i.e. $\pm2\sqrt{2}$, and the solid blue line represent the bound $\pm2$ given by local realistic theories. One notices that even for long times a violation can be found, though the short lived component can no longer directly be measured.
}\label{BIfig}
\end{figure}

\begin{figure}
\centering
(a)\includegraphics[scale=0.9]{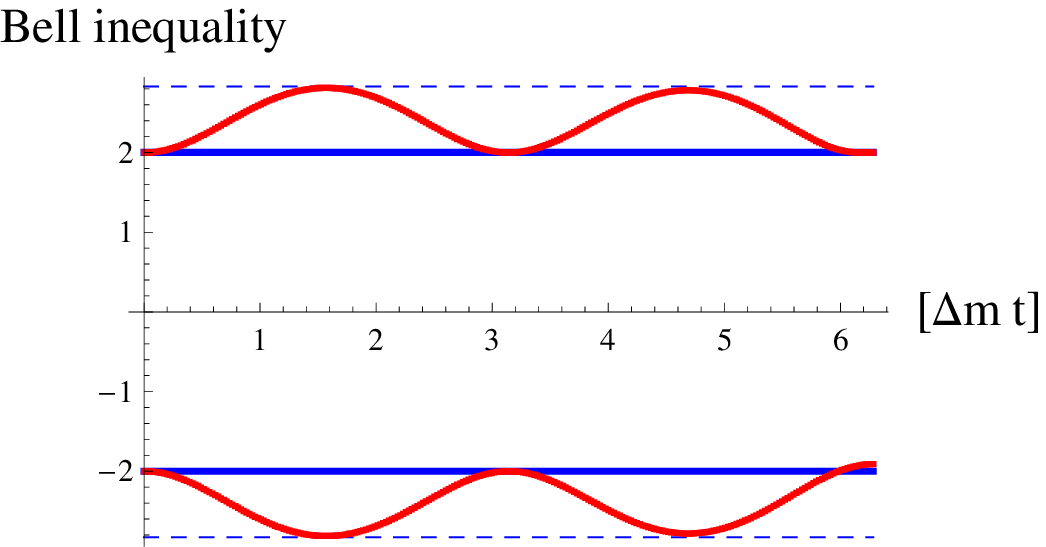}
(b)\includegraphics[scale=0.9]{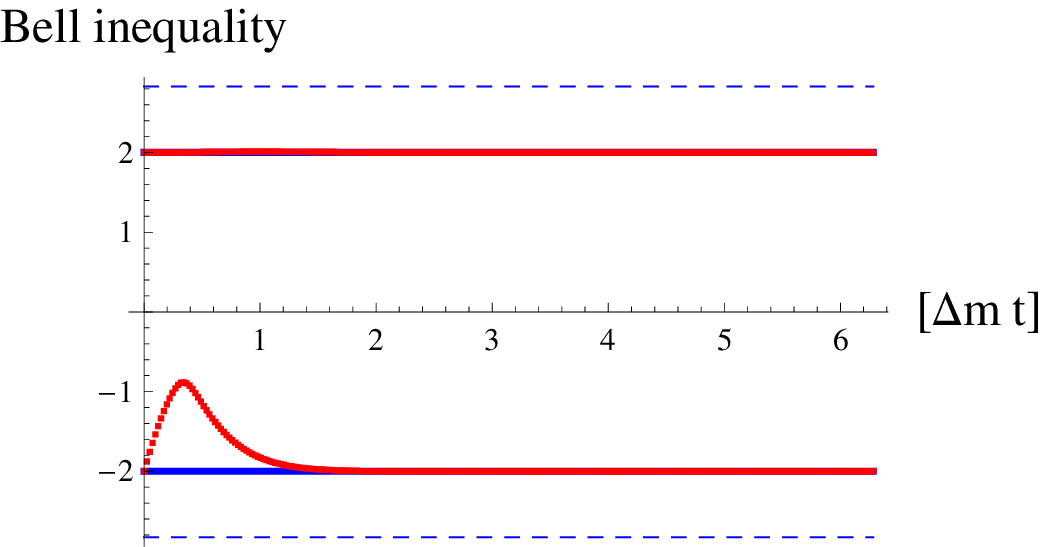}
\caption{(Color online) The maximal violations of the Bell inequality, i.e. the maximal and minimal eigenvalues of the operator (\ref{Belloperator}) for flavor questions for time choices $\{t_n=0,t_m=t,t_{n'}=t,t_{m'}=0\}$ or $\{t_n=t,t_m=0,t_{n'}=0,t_{m'}=t\}$ for (a) $\Gamma_S=\Gamma_L$ and (b) $\Gamma_1=\Gamma_2=1/0.776$ are plotted (red big dots). Here (a) shows the violation if both mass eigenstates wer
e long lived in the neutral kaon system and (b) the values for the B--meson system.
}\label{BIfig2}
\end{figure}

Given our effective operator formalism we can answer the question how much nonlocality is there for the given Bell setting if we vary the times. In Fig.~\ref{BIfig} we plotted the eigenvalues of the Bell operator for different choices corresponding to the maximal/minmal value of the Bell inequality as well as the uncertainty. We find only a small amount of violation (about $2.1$) but huge time regions of possible violations. Moreover, we notice an asymmetric behaviour of the minimal and maximal eigenvalues of the Bell operator which is due to the two different decay constants, as also plotted in Fig.~\ref{BIfig2} for a slow decaying system and for the B--meson system.

In Ref.~~\cite{BellKomm} the authors showed that the maximal violation of the CHSH--Bell inequality is reached when the two operators in the sum of the Bell operator, Eq.~(\ref{Belloperator}), commute. This fact the authors used to construct other relevant Bell inequalities for two--qubit systems. For unstable systems we do not see a one--to--one correspondence between the uncertainty of the summands in the Bell operator and the amount of violation, moreover, $O^{eff}_m\pm O^{eff}_{m'}$ does not necessary describe an observable obtainable in a single measurement.

\section{Summary and Conclusions}

We studied the phenomenology of decaying two--state systems and discussed quantum features from an information theoretic view. For that we developed an effective formalism which allows to handle unstable two-state systems with the usual well developed formalism in Quantum Information Theory. We applied it to the neutral kaon system including the $\mathcal{CP}$ violation, the observed imbalance between matter and antimatter in our universe.

We presented the effective operator in decomposition of the Pauli-matrices and the unity, which shows the complicated change of the Bloch vector in time. The spectral decomposition shows that only one eigenvalue depends on measurement settings and that the corresponding eigenvectors can be interpreted as quasispins evolving in (forward and backward) time normalized to surviving pairs. The second eigenvalue is always $-1$, i.e. it does not depend on the chosen measurement settings. This expresses the fact that we are only interested in quantum features intrinsic to neutral kaons and not about the properties of the different decay channels.

The lower bound on the binary entropies of two chosen observables is given by maximal overlap of the eigenvectors of both observables and encodes the limitations on the available information obtainable by the chosen observables. To obtain this Heisenberg uncertainty in time for meson-antimeson systems we compared measurement settings at time $t=0$ to the same measurement settings at a later time $t$. We find for flavor measurements that the uncertainty becomes maximal for times which are odd multiples of $\frac{\pi}{2}$, while for times which are multiples of $\pi$ only in the case both decay rates are equal the uncertainty becomes zero again as it is the case for non--decaying systems. For considerably different decay constants as in the neutral kaon system the uncertainty never vanishes for any later time measurement, i.e. introducing an persisting lack of information; this is depicted in Fig.~\ref{uncertainty1}.

Due to imbalance of matter and antimatter we derived a maximal uncertainty for short lived measurements at a ``complementary'' time depending on the precise values of the $\mathcal{CP}$ violating parameter $\delta$. This ``complementary'' time is more than twice the time of the time duration for which the probability to misidentify a long lived state as a short lived state or vice versa is equal. In case of long lived measurements the lower bound on the uncertainty relation is constant (about the amount of the  $\mathcal{CP}$ violating parameter). This is illustrated in Fig.~\ref{uncertainty2} and shows the effect of indirect $\mathcal{CP}$ violation on the states persisting their nature in the time evolution.

Then we proceed to entangled bipartite systems. The effective observables simply generalize for multipartite systems by the usual tensor product
structure which is a clear advantage to the open quantum approach. The uncertainty for bipartite systems is straightforwardly obtained as it is the maximum of the product of the scalar products of the eigenvectors of the single effective operator.

Due to the developed effective formalism Bell inequalities, i.e. inequalities deciding whether a local realistic view for kaons is valid, can be formulated in a mathematically more simple form, i.e. as a witness operator. Herewith, we do not need to optimize over the state space parameters and the four different measurement settings, but can simply compute the eigenvalues of the Bell operator to obtain the maximal possible value given by the quantum theory. In case of strangeness measurements we find that the violation is not big, but can be obtained for long time regions. Indeed, also for times when the short lived component has already died out for all practical purposes, i.e. no oscillation can be seen, but since the probability is still nonzero, non-negligible contributions in the Bell operator exist.

We believe with this information theoretic view on unstable two--state systems and, in particular, on the meson-antimeson systems in high energy physics we could enlighten the quantum features in these massive systems and, in particular, the threefold role of time, being responsible for strangeness oscillations, oscillation due to  $\mathcal{CP}$ violation and characterizing the decay property.

\vspace{0.5cm}
\noindent\emph{Acknowledgements:}
Marcus Huber and Christoph Spengler gratefully acknowledge the Austrian Fund project FWF-P21947N16. Andreas Gabriel is supported by the University of Vienna's research grant. Beatrix C. Hiesmayr acknowledges the EU project QESSENCE and wants to thank in particular Reinhold A. Bertlmann for introducing her to the world of neutral kaons and wants to devote this paper to him.

\end{document}